\documentclass[journal]{IEEEtran}
\usepackage{graphicx}
\usepackage{subfigure}
\newcommand{\tabincell}[2]{\begin{tabular}{@{}#1@{}}#2\end{tabular}}
\usepackage{multirow}
\usepackage{amsmath,amssymb,amsfonts}
\usepackage{cite}
\usepackage{algpseudocode}
\usepackage{algorithm}
\usepackage{algorithmicx}
\newtheorem{theorem}{\textbf{Theorem}}  
\newtheorem{lemma}{\textbf{Lemma}} 
\usepackage{xcolor}
\usepackage{textcomp}
\begin{document}
\title{Fast List Decoders for Polarization-Adjusted Convolutional (PAC) Codes}
\author{
	Hongfei~Zhu, Zhiwei~Cao, Yuping~Zhao, Dou~Li and Yanjun~Yang
	
	\thanks{The authors are with the School of Electronics Engineering and Computer Science,
		Peking University, Beijing 100871, China (e-mail: zhuhongfei@pku.edu.cn; cao\_zhiwei@pku.edu.cn; yuping.zhao@pku.edu.cn; lidou@pku.edu.cn; yangyj@pku.edu.cn).}
}
\maketitle

\begin{abstract}
	A latest coding scheme named polarization-adjusted convolutional (PAC) codes is shown to approach the dispersion bound for the code (128,64) under list decoding. However, to achieve the near-bound  performance, the list size of list decoding needs to be excessively large, which leads to insufferable latency. In this paper, to improve the speed of list decoding, fast list decoders for PAC codes are proposed. We define four types of constituent nodes and provide fast list decoding algorithms for each of them. Simulation results present that fast list decoding with three types of constituent nodes can yield exactly the same error-correction performance as list decoding, and reduce more than 50\% time steps for the code (128,64). Moreover, fast list decoding with four types of constituent nodes can further reduce decoding latency with negligible performance degradation.
\end{abstract}
\IEEEpeerreviewmaketitle

\section{Introduction}
\label{Section_Introduction}

\IEEEPARstart{P}{olar} codes are the first family of error-correcting codes with provable capacity-achieving property for systematic binary-input discrete memoryless channels (BI-DMCs) \cite{Polar_Arikan}. The successive cancellation (SC) decoding is one of the most common decoding algorithms of polar codes with a low complexity $O(N\log_2 N)$, where $N$ is the code length. However, there are two main drawbacks associated with SC. Firstly, polar codes decoded with SC  achieve the channel capacity only when $N$ approaches infinity. For practical polar codes of short to moderate lengths, SC falls short in providing satisfactory error-correction performance. Secondly, the serial decoding nature of SC results in low throughput and high latency, and one codeword consumes $2N-2$ time steps without resource constraints \cite{SC_delay}. 

In order to reduce the performance gap between SC and maximum likelihood (ML), the SC List (SCL) decoding algorithm was introduced in \cite{SCL_TalVardy}. Instead of focusing on a single
candidate codeword like SC, the $L$ most probable candidate codewords are allowed to survive concurrently. Under SCL decoding, it was shown that polar codes concatenated with a cyclic redundancy check (CRC) can outperform low-density parity-check (LDPC) codes of the similar length. While SCL provides better error-correction performance than SC, it comes at the cost of lower throughput and higher latency, requiring $2N-2+K$ time steps to be completed, where $K$ is the number of information bits \cite{SCL_delay}. 

Various attempts have been made to improve the speed of SCL decoding. Among these methods, tree pruning was shown to be extremely effective. Four types of constituent nodes, namely rate zero (Rate-0) node with all frozen bits, rate one (Rate-1) node with all information bits, repetition (Rep) node with a single information bit in the most reliable position, and single parity-check (SPC) node with a single frozen bit in the least reliable position, were capable of being decoded in parallel with low-complexity decoding algorithms. Based on the four special nodes, simplified SCL (SSCL) \cite{SSCL} and SSCL-SPC \cite{SSCL-SPC} increased the throughput and reduced the latency significantly with negligible error-correction performance loss with respect to SCL. Nonetheless, SSCL and SSCL-SPC failed to address the redundant path split associated with a specific list size. Therefore, Fast-SSCL \cite{Fast-SSCL} and Fast-SSCL-SPC \cite{Fast-SSCL-SPC} provided a more efficient algorithm for Rate-1 and SPC nodes, respectively. These state-of-the-art algorithms can further reduce the number of path forks and guarantee the error-correction performance preservation.

Recently in \cite{PAC_Arikan}, a new coding scheme named polarization-adjusted convolutional (PAC) codes were proposed. They rely on the concatenation a convolutional transform \cite{ConvolutionalCodes} with the polarization transform. The message vector $\mathbf{v}$ is first encoded using a one-to-one convolutional operation and then transmitted over polarized synthetic channels. Remarkably, under Fano sequential decoding \cite{Fano}, the performance of PAC codes with length 128 and rate 1/2 can reach the finite-length capacity bound  a.k.a. dispersion bound \cite{DispersionBound}. Later in \cite{PAC_FanoVsList} and \cite{PAC_List}, the authors independently proposed the list decoding for PAC codes. Compared with sequential decoding, list decoding is a non-backtracking tree search approach with a fixed complexity. Moreover, it is certainly advantageous in terms of worst-case complexity and at low signal-to-noise ratios (SNRs). However, to achieve the same performance as sequential decoding, the list size $L$ of list decoding needs to be excessively large ($L \geq 256$), leading to insufferable latency. 

In this paper, to alleviate the tremendous latency of list decoding, we propose fast list decoders for PAC codes. Four types of constituent nodes are defined based on the message vector $\mathbf{v}$, namely Rate-0 nodes, Rate-1 nodes, Rev (reversal) nodes and SPC nodes. For each constituent node, its fast list decoding algorithm is provided along with necessary proofs. Numerical results illustrate that fast list decoding with three types of constituent nodes, i.e., Rate-0, Rate-1 and Rev nodes provides an exact match to list decoding with no error-correction performance loss. In addition, the number of time steps can be significantly reduced. Moreover, the introduction of SPC nodes further improves the latency and results in neglectable performance degradation.

The remainder of this paper is organized as follows: Section \ref{Section_Background} provides a background on PAC codes and list  decoding. Section \ref{Section_ProposedAlgorithms} introduces the proposed fast list decoders for PAC codes. Simulation results are provided in Section \ref{Section_SimulationResults}. Finally, Section \ref{Section_Conclusion} draws the main conclusions of the paper.

\section{Background}
\label{Section_Background}

\subsection{PAC Codes}
\label{Subsection_PACcodes}
\begin{figure}[t!]
	\centering
	\includegraphics[width=.45\textwidth]{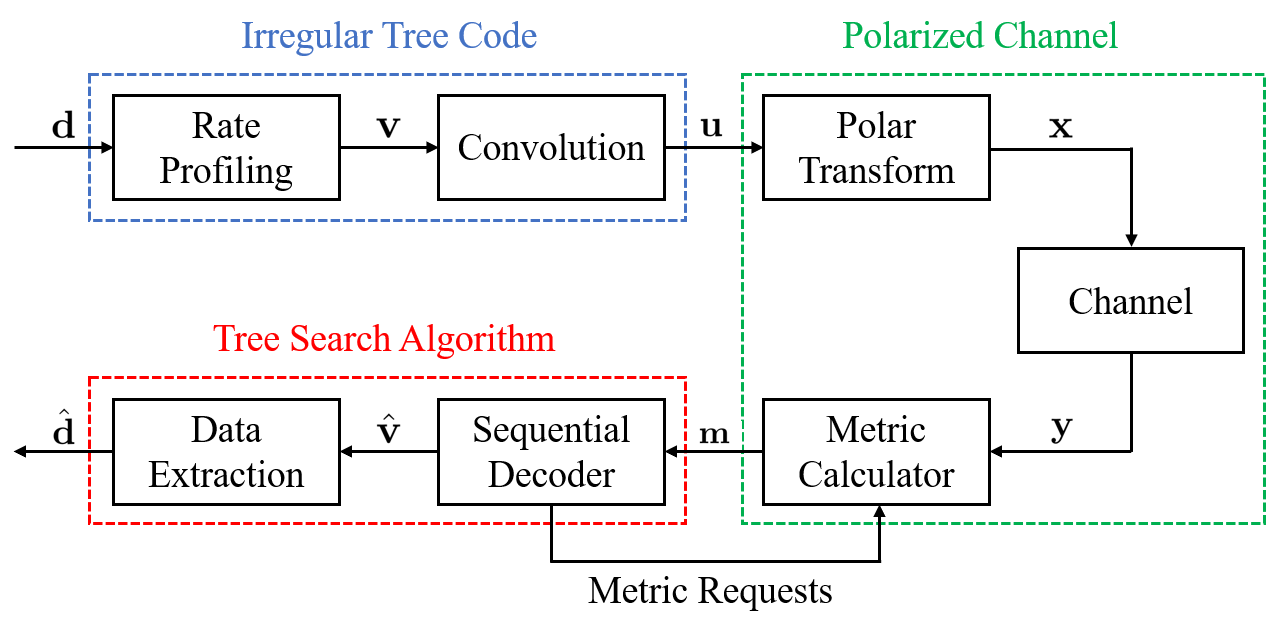}
	\caption{PAC coding scheme.} 
	\label{Figure_BlockDiagram} 
\end{figure}
The PAC coding scheme is shown in Fig. \ref{Figure_BlockDiagram}. A PAC code is specified by four parameters
$(N, K, \mathcal{A}, \mathbf{c})$, where $N$ is the codeword length, $K$ is the length of information bits, $\mathcal{A}$ is the index set of information bits, and $\mathbf{c}$ is the impulse response of convolution. The codeword length is constrained to be a power of two, i.e., $N =2^n$ for $n \geq 1$, just as polar codes. The code rate is denoted as $R=K/N$. 

At the transmitter, a rate-profiling block inserts the information vector $\mathbf{d}=(d_0,d_1,...,d_{K-1})$ into a message vector $\mathbf{v}=(v_0,v_1,...,v_{N-1})$ in accordance with a information bits set $\mathcal{A}$ so that $\mathbf{v}_\mathcal{A} = \mathbf{d}$ and $\mathbf{v}_{\mathcal{A}^c} = \mathbf{0}$. Then $\mathbf{u}=(u_0,u_1,...,u_{N-1})$ can be computed as $\mathbf{u} = \mathbf{v}\mathbf{T}$ where $\mathbf{T}$ is an upper-triangular Toeplitz matrix denoting an one-to-one convolution operation. The PAC codeword $\mathbf{x}=(x_0,x_1,...,x_{N-1})$ is obtained from $\mathbf{u}$ by $\mathbf{x} = \mathbf{u}\mathbf{P}_n$ where $\mathbf{P}_n$ is the polar transform defined as the $n$-th Kronecker power of $\mathbf{P}=[1,0;1,1]$ , i.e., $\mathbf{P}_n=\mathbf{P}^{\otimes n}$ ($n=\log_2N$). As usual, we characterize the convolution operation by an impulse response $\mathbf{c} = (c_0,c_1,...,c_m)$, where by convention we assume that $c_0= 1$ and $c_{m}= 1$. The number of states of the convolutional shift register is $2^m$ and the parameter $m + 1$ is called the constraint length of the convolution. 

The vector $\mathbf{x}$  is transmitted through a noisy channel and received as the vector $\mathbf{y}$. At the receiver, Fano sequential decoding \cite{Fano} of the underlying convolutional code can be employed to decode the data carrier vector $\mathbf{v}$. Notably, the path metrics at the input to the sequential decoder are obtained via repeated calls to the SC decoder for the underlying polar code. The performance of PAC codes is more sensitive to the choice of $\mathcal{A}$ than to $\mathbf{c}$, and the RM rule \cite{RM_1,RM_2} was shown to be the best method to  design $\mathcal{A}$ in terms of frame error rate (FER) performance \cite{PAC_Arikan}.

\subsection{List Decoding of PAC Codes}
Sequential decoding is a backtracking search algorithm with unfixed complexity varying with SNRs. Representing PAC codes as the special case of polar codes with dynamically frozen bits \cite{DynamicFrozenBits}, the authors in \cite{PAC_FanoVsList} and  \cite{PAC_List} independently put forward a non-backtracking search algorithm with fixed complexity named list decoding  for PAC codes.
Since PAC codes combine convolution and polar mapping, the list decoding process of PAC codes consists of two trees: code tree for convolutional codes and binary tree for polar codes. In the following, we take \textit{PAC}$(8,4,\{3,5,6,7\},(1,1,1))$ for example.
\subsubsection{Code Tree for Convolutional Codes}
\begin{figure}[t!]
	\centering
	\includegraphics[width=.2\textwidth]{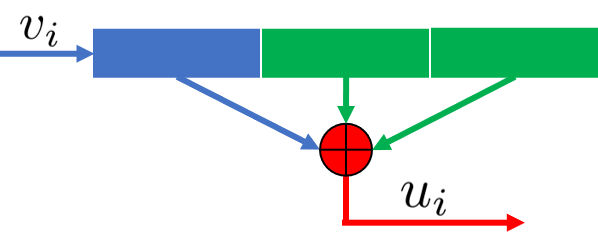}
	\caption{The shift register corresponding to $\mathbf{c}=(1,1,1)$.} 
	\label{Figure_Register} 
\end{figure}
\begin{figure}[t!]
	\centering
	\includegraphics[width=.3\textwidth]{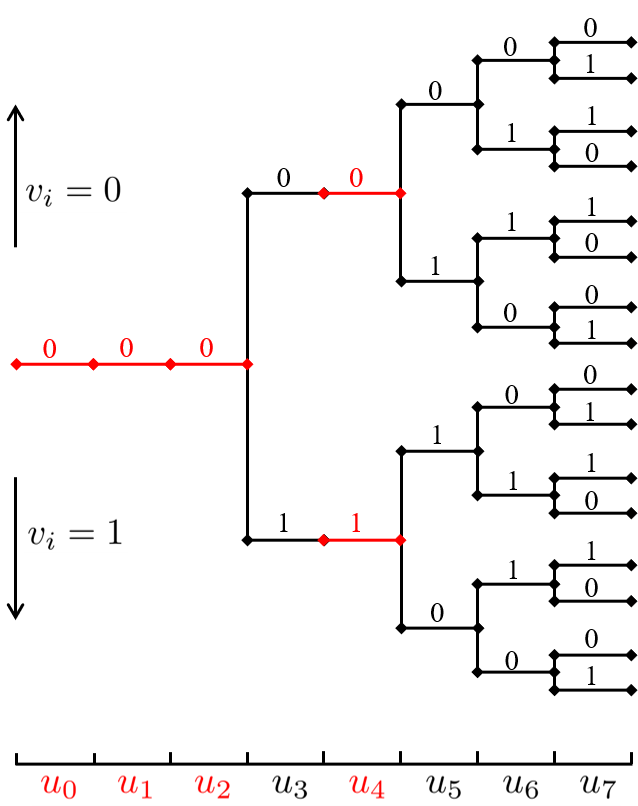}
	\caption{Irregular code tree of the rate-1 convolutional code inside the PAC code.} 
	\label{Figure_CodeTree} 
\end{figure}
Fig. \ref{Figure_Register} displays the shift register corresponding to $\mathbf{c}=(1,1,1)$.
Note that the state of the register resets before each codeword enters. Under the frozen-bit constraint imposed by rate-profiling, the rate-1 convolutional code yields an irregular code tree, which is illustrated in Fig. \ref{Figure_CodeTree}. The code tree branches only when there is a new information bit $v_i, i \in \mathcal{A}$ going into the register. When there is branching in the code tree at some stage $i \in \mathcal{A}$,  by convention, the upper branch corresponds to $v_i = 0$ and the lower branch corresponds to $v_i = 1$. Other nodes of the code tree are in one-to-one correspondence with the convolution input words $\mathbf{v}$ satisfying the constraint $\mathbf{v}_{\mathcal{A}^c} = 0$.

\subsubsection{Binary Tree for Polar Codes}
\label{Subsubsection_BinaryTree}
\begin{figure}[t!]
	\centering
	\includegraphics[width=.4\textwidth]{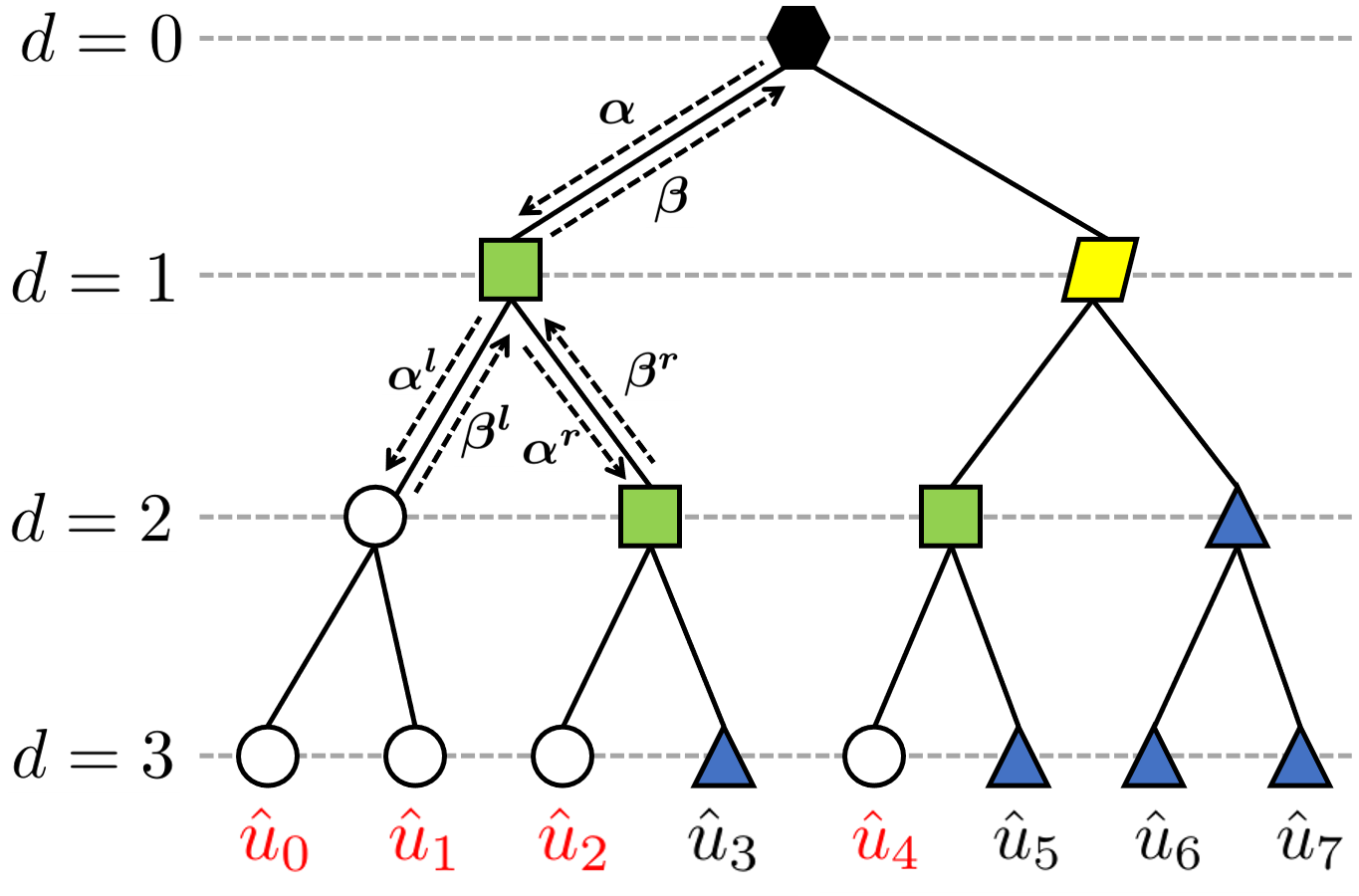}
	\caption{Binary tree of the polar code inside the PAC code. White circles are Rate-0 nodes, blue triangles are Rate-1 nodes, green squares are Rev nodes, yellow parallelograms are SPC nodes and black hexagons are general nodes.} 
	\label{Figure_BinaryTree} 
\end{figure}
The binary tree of the polar code inside the PAC code is presented in Fig. \ref{Figure_BinaryTree}. 
Given a node $o$  of width $N_o=2^{n-d_o}$ at depth $d_o$ in the binary tree, we use $\mathcal{V}_o$ to denote the set of nodes of the subtree rooted
at node $o$. We let $I(o)$ denote the index of a leaf node $o$. Furthermore, to each
node $o$ we associate the set
\begin{align}
\mathcal{I}_o=\{I(q): q \in \mathcal{V}_o \ {\rm and} \ q \ {\rm is} \ {\rm a} \ {\rm leaf} \ {\rm node}\}
\end{align}
containing the indices of all leaf nodes that are descendents of node $o$. Unless specified, we use $\boldsymbol{\alpha}=[\alpha_0, . . . , \alpha_{N_o-1}]$, $\boldsymbol{\beta}= [\beta_0, . . . , \beta_{N_o-1}]$, $\mathbf{u}=[u_0, . . . , u_{N_o-1}]$, and $\mathbf{v}=[v_0,...,v_{N_o-1}]$ to denote the log-likelihood ratio (LLR) vector on the top of the subtree, the bit vector on the top of the subtree, the bit vector at the bottom of the subtree, and the message vector corresponding to $\mathbf{u}$, respectively. Additionally,  $\mathbf{P}_{n-d_o}$ is the $(n-{d_o})$-th Kronecker power of $\mathbf{P}=[1,0;1,1]$ and $\mathbf{s}$ is the state vector of the shift register consisting of $m$ bits. Note that the inverse of $\mathbf{P}_{n-d_o}$ is itself \cite{Pn_Property}. 

The SC decoding process for the polar code can be depicted as follows. The vertex node at depth $d = 0$ of the decoding tree is fed with the LLR values received from the channel. Subsequently for each node $o$, the soft LLR values in $\boldsymbol{\alpha}$ are passed from a parent node to its child nodes and the hard bits in $\boldsymbol{\beta}$ are passed from the opposite direction. The elements of the left messages $\boldsymbol{\alpha^l} = [\alpha_0^l,  . . . , \alpha_{\frac{N_o}{2}-1}^l]$ and right messages $\boldsymbol{\alpha^r} = [\alpha_0^r, . . . , \alpha_{\frac{N_o}{2}-1}^r]$ are calculated as 
\begin{align}
\label{Equation_f}
\alpha_i^l&=\text{sgn}(\alpha_i) \cdot \text{sgn}(\alpha_{i + \frac{N_o}{2}}) \cdot \min(|\alpha_i|,|\alpha_{i + \frac{N_o}{2}}|) \\
\label{Equation_g}
\alpha_i^r&=(1-2\beta_i^l) \cdot \alpha_i + \alpha_{i + \frac{N_o}{2}};
\end{align}
where
\begin{align}
	\text{sgn}(x)=\begin{cases}
	1, &x>0 \\
	0, &x=0 \\
	-1, &x<0
	\end{cases}
\end{align}
The elements of $\boldsymbol{\beta}$ are computed using the left messages  $\boldsymbol{\beta}^l = [\beta_0^l,. . . , \beta_{\frac{N_o}{2}-1}^l]$ and right messages  $\boldsymbol{\beta}^r = [\beta_0^r,  . . . , \beta_{\frac{N_o}{2}-1}^r]$
\begin{align}
\beta_i=\begin{cases}
\beta_i^l \oplus \beta_i^r,  &\text{if} \ i < \frac{N_o}{2}      \\
\beta_{i-\frac{N_o}{2}}^r,   &\text{otherwise}
\end{cases}
\end{align}

\subsubsection{Path Split}
\label{Section_PathSplit}
At the beginning of list decoding, there is a single path in the list. A path metric (PM) is associated to each path and the process of path split is described as follows:
\begin{itemize}
	\item When the index of the current bit $v_i$ is in the set $\mathcal{A}^c$, the decoder knows its value, usually $v_i = 0$ and therefore it is encoded into $u_i$ based on the impulse response $\mathbf{c}$ and the current state $\mathbf{s}$ of the shift register. Then the PM of the $i$-th bit for the $l$-th path is computed as 
	\begin{align}
	\text{PM}_{i_l}=\begin{cases}
	\text{PM}_{{i-1}_l},                  &\text{if}  \ \hat{u}_{i_l} = \frac{1}{2} (1-\text{sgn}(\alpha_{i_l}))  \\
	\text{PM}_{{i-1}_l} + |\alpha_{i_l}|, &\text{otherwise}
	\end{cases}
	\label{Equation_PM}
	\end{align}
	Whereafter the shift register of the $l$-th path is updated and $u_i$ is fed back into the SC decoder as mentioned in Section \ref{Subsubsection_BinaryTree}. 
	\item If the index of the current bit $v_i$ is in the set $\mathcal{A}$, there are two options for $v_i$, i.e.,  $v_i =  0$ and $1$. For each option of $0$ and $1$, the aforementioned process for $i \in \mathcal{A}^c$ including convolutional transformation, state update and PM calculation is performed and then the two transformed values $u_i = 0$ and $1$ are fed back into SC process.	We can see that at every estimation for information bits, the number of paths doubles. Hence, in order to limit the increase in the complexity, only the $L$ paths with lowest PMs survive while others are discarded.
\end{itemize}

Simulation results demonstrate that list decoding has distinct advantages over sequential decoding in certain scenarios such as low-SNR regimes or situations where the worst-case complexity/latency is the primary constraint \cite{PAC_List}. Furthermore, PAC codes using list decoding can also approach dispersion bound of length 128 and rate 1/2, just as sequential decoding \cite{PAC_FanoVsList,PAC_List}. However, the near-bound performance of list decoding requires extremely large list size $(L \geq 256)$, leading to considerable latency which is not feasible in practice. Therefore, based on Fast-SSCL-SPC algorithm for polar codes \cite{Fast-SSCL-SPC}, we propose fast list decoders for PAC codes as presented in the following.

\section{Fast List Decoders for PAC Codes}
\label{Section_ProposedAlgorithms}
It is known that for polar codes, PMs rely on the LLR values at the top of the binary tree identified by their constituent nodes: thus, it is not necessary to traverse the tree to compute them \cite{SSCL-SPC}. In this section, we define four types of constituent nodes for PAC codes based on the message vector $\mathbf{v}$, namely Rate-0 nodes, Rate-1 nodes, Rev nodes and SPC nodes. For each constituent node, its fast list decoding algorithm is provided along with necessary proofs. Subroutines inside all algorithms are provided in Appendix \ref{Appendix_Subroutines}. Eventually, compared with list decoding, time steps of fast list decoding for each constituent node are analyzed.

\subsection{Rate-0 Nodes}
We say that a node $o$ is a Rate-0 node with respect to $\mathcal{A}^c$ if $\mathcal{I}_o \in \mathcal{A}^c$, i.e., if all bits in $\mathbf{v}$ are frozen bits. For each path, when SC decoding reaches a rate-0 node $o$ plotted in white circles in Fig. \ref{Figure_BinaryTree}, it can immediately set $\mathbf{v}=\mathbf{0}$ without utilizing  $\boldsymbol{\alpha}$ to activate its children. Then $\mathbf{u}$ can be obtained through convolutional encoding and $\boldsymbol{\beta}$ can be computed using polar encoding. Finally, PMs are updated using $\boldsymbol{\alpha}$ and $\boldsymbol{\beta}$ following the same formula as (\ref{Equation_PM}). Notice that no path split is needed for Rate-0 nodes. Algorithm \ref{Algorithm_Rate0} illustrates the concrete steps.

\begin{algorithm}[t]	
	\caption{Fast List Decoding of a Rate-0 Node $o$} 
	\hspace*{0.02in} {\bf Input:} 
	$\boldsymbol{\alpha}[l], \mathbf{s}[l], PM_l, l=0,...,L-1$, $\mathbf{P}_{n-d_o},\mathbf{c}$\\
	\hspace*{0.02in} {\bf Output:} 
	$\boldsymbol{\beta}[l], \mathbf{s}[l], PM_l, \mathbf{\hat{v}}[l], \mathbf{\hat{u}}[l], l=0,...,L-1$
	\begin{algorithmic}[1]
		\For{$l \leftarrow 0$ \textbf{to} $L-1$}
			\For{$i \leftarrow 0$ \textbf{to} $N_o-1$}
				\State $\hat{v}_i[l] \leftarrow 0$
				\State \texttt{// convolutional encoding}
				\State [$\hat{u}_i[l],\mathbf{s}[l]$] $\leftarrow$ conv1bitEnc($\hat{v}_i[l], \mathbf{s}[l],\mathbf{c}$)
			\EndFor
			\State $\boldsymbol{\beta}[l]	\leftarrow \hat{\mathbf{u}}[l] \cdot \mathbf{P}_{n-d_o}	$ \texttt{// polar encoding}
			\State $PM_l \leftarrow $ calcPM($PM_l,\boldsymbol{\alpha}[l],\boldsymbol{\beta}[l], N_o$)		
		\EndFor		
	\end{algorithmic}
\label{Algorithm_Rate0}
\end{algorithm}

\subsection{Rate-1 Nodes}
We say that a node $o$ is a Rate-1 node with respect to $\mathcal{A}$ if $\mathcal{I}_o \in \mathcal{A}$, i.e., if all bits in $\mathbf{v}$ are  information bits. In Fig. \ref{Figure_BinaryTree}, Rate-1 nodes are plotted in blue triangles. The process of path split can be carried out on the top of the subtree following the same rule as illustrated in Section \ref{Section_PathSplit}. Although a Rate-1 node $o$ with width $N_o$ contains $N_o$ information bits, the minimum number of path splits  incurring no error-correction performance loss is $\min(L-1,N_o)$ based on the sorted LLR vector $\boldsymbol{\tilde{\alpha}}$ \cite{Fast-SSCL}. Other bits in $\boldsymbol{\beta}$ can be directly judged by their corresponding LLRs in  $\boldsymbol{\alpha}$. After $\boldsymbol{\beta}$ is obtained, $\mathbf{u}$ can be calculated through the inverse process of polar encoding and subsequently  $\mathbf{v}$ can be calculated through the inverse process of convolutional encoding. We give the pseudo-code in Algorithm \ref{Algorithm_Rate1}. 
\begin{algorithm}[t]	
	\caption{Fast List Decoding of a Rate-1 Node $o$} 
	\hspace*{0.02in} {\bf Input:} 
	$\boldsymbol{\alpha}[l], \mathbf{s}[l], PM_l, l=0,...,L-1$, $\mathbf{P}_{n-d_o},\mathbf{c}$\\
	\hspace*{0.02in} {\bf Output:} 
	$\boldsymbol{\beta}[l], \mathbf{s}[l], PM_l, \mathbf{\hat{v}}[l], \mathbf{\hat{u}}[l], l=0,...,L-1$
	\begin{algorithmic}[1]		
		\For{$l \leftarrow 0$ \textbf{to} $L-1$}  \texttt{// judgment and sorting}
			\State $\boldsymbol{\beta}[l] \leftarrow \frac{1}{2}(1-$sgn($\boldsymbol{\alpha}[l]$)
			\State [$|\boldsymbol{\tilde{\alpha}}[l]|,\mathbf{\tilde{I}}[l]$] $\leftarrow $ sort($|\boldsymbol{\alpha}[l]|$,`ascend')			
		\EndFor
		\For{$i \leftarrow 0$ \textbf{to} $\min(L-1,N_o)-1$}   \texttt{// path split}
			\For{$l \leftarrow 0$ \textbf{to} $L-1$}
				\State $ll \leftarrow l+L$		
				\State [$\boldsymbol{\beta'}[l],\boldsymbol{\beta'}[ll]$]  $\leftarrow$ [$\boldsymbol{\beta}[l],\boldsymbol{\beta}[l]$]
				\State [$\mathbf{s'}[l],\mathbf{s'}[ll]$] $\leftarrow$ [$\mathbf{s}[l],\mathbf{s}[l]$]
				\State [$\boldsymbol{\tilde{\alpha}'}[l],\boldsymbol{\tilde{\alpha}'}[ll] $ $\leftarrow$
					   [$\boldsymbol{\tilde{\alpha}}[l],\boldsymbol{\tilde{\alpha}}[l]$]
		        \State [$\mathbf{\tilde{I}'}[l],\mathbf{\tilde{I}'}[ll] $] $\leftarrow$
				[$\mathbf{\tilde{I}}[l],\mathbf{\tilde{I}}[l] $]
				\State [$\tilde{\beta}'_i[l],\tilde{\beta}'_i[ll]$] $\leftarrow$ [$0,1$]
				\State $PM'_l \leftarrow $ calcPM($PM_l,\tilde{\alpha}'_i[l],\tilde{\beta}'_i[l], 1$)	
				\State $PM'_{ll} \leftarrow $ calcPM($PM_{l},\tilde{\alpha}'_i[ll],\tilde{\beta}'_i[ll], 1$)
			\EndFor	
			\State [$\mathbf{\widetilde{PM}},\mathbf{\tilde{l}}$] $\leftarrow$ sort([$PM'_0,...,PM'_{2L-1}$],`ascend')
			\For {$l \leftarrow 0$ \textbf{to} $L-1$}
				\State $PM_l$ $\leftarrow$ $\widetilde{PM}_l$ 
				\State $\mathbf{\tilde{I}}[l]$ $\leftarrow$ $\mathbf{\tilde{I}'}[\tilde{l}_l]$
				\State $\boldsymbol{\tilde{\alpha}}[l]$ $\leftarrow$ $\boldsymbol{\tilde{\alpha}'}[\tilde{l}_l]$
				\State $\mathbf{s}[l]$ $\leftarrow$ $\mathbf{s'}[\tilde{l}_l]$					
				\State $\boldsymbol{\beta}[l]$ $\leftarrow$ $\boldsymbol{\beta'}[\tilde{l}_l]$			
				\State $\beta_{\tilde{I}_i[l]}[l]$ $\leftarrow$ $\tilde{\beta}'_i[\tilde{l}_l]$				
			\EndFor			 
		\EndFor
		\For{$l \leftarrow 0$ \textbf{to} $L-1$}
			\State \texttt{// inverse polar encoding}
			\State $\hat{\mathbf{u}}[l]	\leftarrow  \boldsymbol{\beta}[l] \cdot \mathbf{P}_{n-d_o}$	 
			\For{$i \leftarrow 0$ \textbf{to} $N_o-1$}	
			    \State \texttt{// inverse convolutional encoding}			
				\State [$\hat{v}_i[l],\mathbf{s}[l]$] $\leftarrow$ conv1bitInvEnc($\hat{u}_i[l], \mathbf{s}[l],\mathbf{c}$)				
			\EndFor
		\EndFor	
	\end{algorithmic}
	\label{Algorithm_Rate1}
\end{algorithm}

\subsection{Rev Nodes}
We say that a node $o$ is a Rev node if the first $N_o-1$ bits in $\mathbf{v}$ are frozen bits while the last bit $v_{N_o-1}$ is an information bit. The Rev nodes are depicted in green squares in Fig. \ref{Figure_BinaryTree}. 

\begin{lemma}
	The elements of the last row of $\mathbf{P}_{n}, n \geq 1$ are all ones.
	\label{Lemma1}
\end{lemma}

\textit{Proof:} We prove this lemma by induction. When $n=1$, we have $\mathbf{P_1}=\mathbf{P}=[1,0;1,1]$, the lemma holds. Supposing that this lemma holds for $n-1, n \geq 2$, then we have 
\begin{equation}
\begin{split}
	\mathbf{P}_{n}&=\mathbf{P} \otimes \mathbf{P}_{n-1}\\
&=
\left[                
\begin{array}{cc}   
1 & 0 \\  
1 & 1 \\  
\end{array}
\right]                
\otimes \mathbf{P}_{n-1}\\
&=					
\left[                
\begin{array}{cc}   
\mathbf{P}_{n-1} & \mathbf{0} \\  
\mathbf{P}_{n-1} & \mathbf{P}_{n-1} \\  
\end{array}
\right]
\end{split}
\label{Equation_Pn1}
\end{equation}

It can be seen that the last row of $\mathbf{P}_{n}$ is constructed by concatenation the last row of $\mathbf{P}_{n-1}$ twice. Since we have assumed that the elements of the last row of $\mathbf{P}_{n-1}$ are all ones,  the elements of last row of $\mathbf{P}_{n}$ are also all ones.
$\hfill\blacksquare$

Based on Lemma \ref{Lemma1}, we can obtain the following theorem for Rev nodes. 
\begin{theorem}
	As for a Rev node $o$, Assuming that $\boldsymbol{\beta}_0$ is the bit vector corresponding to $v_{N_o-1}=0$ while $\boldsymbol{\beta}_1$ corresponds to $v_{N_o-1}=1$, then we have 
	\begin{align}
	\boldsymbol{\beta}_0 \oplus \boldsymbol{\beta}_1 = \mathbf{1}
	\end{align}	
	\label{Theorem1}
\end{theorem}

\textit{Proof:} 
We have known that the first $N_o-1$ bits in $\mathbf{v}$ are fixed to zeros, i.e., $v_i=0,i=0,...,N_o-2$ while only the last bit have two values, i.e., $v_{N_o-1}= 0$ or $1$. Moreover, in Section \ref{Subsection_PACcodes} we have assumed that the first element in $\mathbf{c}$ is $1$, i.e., $c_0=1$. Therefore, after $N_o$-bit convolutional encoding, two cases of $\mathbf{u}$ will only differ in the last bit $u_{N_o-1}$ while the first $N_o-1$ bits are exactly same. In addition, according to Lemma \ref{Lemma1}, the last row of $\mathbf{P}_{n-d_o}$ are all ones. Consequently, when polar encoding $\boldsymbol{\beta}	= \mathbf{u} \cdot \mathbf{P}_{n-d_o}$ is implemented, $u_{N_o-1}$ will participate in the computation of every element in $\boldsymbol{\beta}$. If the value of $u_{N_o-1}$ is reversed, every element in $\boldsymbol{\beta}$ will also be reversed, which leads to $\boldsymbol{\beta}_0 \oplus \boldsymbol{\beta}_1 = \mathbf{1}$.   
$\hfill\blacksquare$

Since only an information bit is included in the Rev node, only one path split is needed for list decoding. And by Theorem \ref{Theorem1}, we can get an efficient algorithm as expressed in Algorithm \ref{Algorithm_Rev}. The $N_o-1$ frozen bits $v_i,i=0,...,N_o-2$ should be set to 0 firstly. The information bit $v_{N_o-1}$ results in the path expansion of each path, and comparing $\boldsymbol{\alpha}$ with $\boldsymbol{\beta}$, we can compute the PMs on the top of the subtree. Notice that Theorem \ref{Theorem1} can simplify the computation of $\boldsymbol{\beta}$. Eventually, after path elimination, we can get $v_{N_o-1}$ of each path.
\begin{algorithm}[t]	
	\caption{Fast List Decoding of a Rev Node $o$} 
	\hspace*{0.02in} {\bf Input:} 
	$\boldsymbol{\alpha}[l], \mathbf{s}[l], PM_l, l=0,...,L-1$, $\mathbf{P}_{n-d_o},\mathbf{c}$\\
	\hspace*{0.02in} {\bf Output:} 
	$\boldsymbol{\beta}[l], \mathbf{s}[l], PM_l, \mathbf{\hat{v}}[l], \mathbf{\hat{u}}[l], l=0,...,L-1$
	\begin{algorithmic}[1]
		\For{$l \leftarrow 0$ \textbf{to} $L-1$}	
			\For{$i \leftarrow 0$ \textbf{to} $N_o-2$}	
				\State $\hat{v}_i[l]$ $\leftarrow$ $0$		
				\State \texttt{// convolutional encoding}	
				\State [$\hat{u}_i[l],\mathbf{s}[l]$] $\leftarrow$ conv1bitEnc($\hat{v}_i[l],\mathbf{s}[l],\mathbf{c}$)		
			\EndFor
		\EndFor
		\State \texttt{// path split}
		\For{$l \leftarrow 0$ \textbf{to} $L-1$}  
			\State $ll \leftarrow l+L$
			\State [$\mathbf{s'}[l],\mathbf{s'}[ll]$] $\leftarrow$ [$\mathbf{s}[l],\mathbf{s}[l]$]
			\State [$\hat{\mathbf{v}}'[l],\hat{\mathbf{v}}'[ll]$] $\leftarrow$ [$\hat{\mathbf{v}}[l],\hat{\mathbf{v}}[l]$]					
			\State [$\hat{\mathbf{u}}'[l],\hat{\mathbf{u}}'[ll]$] $\leftarrow$ [$\hat{\mathbf{u}}[l],\hat{\mathbf{u}}[l]$]		
			\State [$\hat{v}'_{N_o-1}[l],\hat{v}'_{N_o-1}[ll]$] $\leftarrow$ [$0,1$]			
			\State [$\hat{u}'_{N_o-1}[l],\mathbf{s'}[l]$] $\leftarrow$ conv1bitEnc($\hat{v}'_{N_o-1}[l],\mathbf{s'}[l],\mathbf{c}$)
			\State [$\hat{u}'_{N_o-1}[ll],\mathbf{s'}[ll]$] $\leftarrow$ conv1bitEnc($\hat{v}'_{N_o-1}[ll],\mathbf{s'}[ll],\mathbf{c}$)			
			\State $\boldsymbol{\beta'}[l]	\leftarrow \hat{\mathbf{u}}'[l] \cdot \mathbf{P}_{n-d_o}	$
			\State $\boldsymbol{\beta'}[ll]	\leftarrow \mathbf{1} - \boldsymbol{\beta'}[l] $
			\State $PM'_l \leftarrow $ calcPM($PM_l,\boldsymbol{\alpha}[l],\boldsymbol{\beta'}[l], N_o$)
			\State $PM'_{ll} \leftarrow $ calcPM($PM_l,\boldsymbol{\alpha}[l],\boldsymbol{\beta'}[ll], N_o$)	
		\EndFor
		\State [$\mathbf{\widetilde{PM}},\mathbf{\tilde{l}}$] $\leftarrow$ sort([$PM'_0,...,PM'_{2L-1}$],`ascend')
		\For {$l \leftarrow 0$ \textbf{to} $L-1$}
			\State $PM_l$ $\leftarrow$ $\widetilde{PM}_l$
			\State $\mathbf{s}[l]$ $\leftarrow$ $\mathbf{s'}[\tilde{l}_l]$
			\State $\boldsymbol{\beta}[l]$ $\leftarrow$ $\boldsymbol{\beta'}[\tilde{l}_l]$
			\State $\mathbf{\hat{u}}[l] \leftarrow \mathbf{\hat{u}'}[\tilde{l}_l]$
			\State $\mathbf{\hat{v}}[l] \leftarrow \mathbf{\hat{v}'}[\tilde{l}_l]$			
		\EndFor			
	\end{algorithmic}
	\label{Algorithm_Rev}
\end{algorithm}

\subsection{SPC Nodes}
We say that a node $o$ is a SPC node if the first  bit $v_0$ in $\mathbf{v}$ is a frozen bit while the last $N_o-1$ bits are information bits. Fig.  \ref{Figure_BinaryTree} depicts the SPC nodes in yellow parallelograms.

\begin{lemma}
	The first row of $\mathbf{P}_{n}, n \geq 1$ contains only one 1 at the first position while other rows contain even number of ones. We can formulate this expression as 
	\begin{equation}
		\begin{split}
		P_n(0,0) &= 1, P_n(0,j) = 0, 1 \leq j \leq 2^n-1\\
		\bigoplus_{j=0}^{2^n-1} P_n(i,j) &= 0, 1 \leq i \leq 2^n-1 \\
		\end{split}
	\label{Equation_Pn2}
	\end{equation}
	\label{Lemma2}
\end{lemma}

\textit{Proof:} We prove this lemma by induction. When $n=1$, we have $\mathbf{P_1}=\mathbf{P}=[1,0;1,1]$, the lemma holds. Then we suppose this lemma holds for $n-1, n \geq 2$. According to (\ref{Equation_Pn1}), 
the first half rows in $\mathbf{P}_{n}$ are constructed by concatenation  $\mathbf{P}_{n-1}$ with $\mathbf{0}$, while the second half rows in $\mathbf{P}_{n}$ are constructed by concatenation  $\mathbf{P}_{n-1}$ twice. Hence, the number of ones of the first half rows in $\mathbf{P}_{n}$ are the same as $\mathbf{P}_{n-1}$ while the number of ones of the second half rows in $\mathbf{P}_{n}$ are even. Since we have assumed that the first row of $\mathbf{P}_{n-1}$ contains only one 1 at the first position while other rows contain even number of ones,  we can get that the first row of $\mathbf{P}_{n}$ contains only one 1 at the first position while other rows contain even number of ones.
$\hfill\blacksquare$

Based on Lemma \ref{Lemma2}, we can obtain the following theorem for SPC nodes. 
\begin{theorem}
	As for a SPC node $o$, the bit vector $\boldsymbol{\beta}$ will satisfy 
	\begin{align}
	\bigoplus_{j=0}^{N_o-1} \beta_j = u_0
	\end{align}
	\label{Theorem2}
\end{theorem}

\textit{Proof:} According to matrix multiplication, we have
\begin{equation}
\begin{split}
\bigoplus_{j=0}^{N_o-1} \beta_j 
&= \bigoplus_{j=0}^{N_o-1} \bigoplus_{i=0}^{N_o-1} u_i   P(i,j)                    \\
&= \bigoplus_{i=0}^{N_o-1} u_i \bigoplus_{j=0}^{N_o-1}    P(i,j)    \\
								&= (u_0\bigoplus_{j=0}^{N_o-1}    P(0,j)) \oplus (\bigoplus_{i=1}^{N_o-1} u_i \bigoplus_{j=0}^{N_o-1}    P(i,j))  \\
								&=u_0\\
\end{split}
\end{equation} 
The last equation is obtained by substituting (\ref{Equation_Pn2}).
$\hfill\blacksquare$

When the list decoder reaches a SPC node $o$, the frozen bit $v_0$ should be  set to 0 and $u_0$ can be immediately obtained by 1-bit convolutional encoding. After the hard decision of $\boldsymbol{\alpha}$, we can use $\boldsymbol{\beta}$ and $u_0$ to get the check value 
\begin{align}
\gamma = \bigoplus_{j=0}^{N_o-1} \beta_j \oplus u_0 
\label{Equation_gamma}
\end{align}

To keep $\gamma=0$, the least reliable bit should first decide whether to reverse and then two bits need to be reversed at the same time. The exact number of path splits to preserve the same performance as list decoding is $\binom{N_o}{2}=N_o(N_o-1)/2$, which is quadratic with $N_o$ and it is more complex than the time required by list decoding which is linear with $N_o$ \cite{SSCL-SPC}. However, an approximate algorithm will limit the number of path splits to $\min(L-1,N_o-1)$, which is so effective that the performance loss is negligible\cite{Fast-SSCL-SPC}. The details of the fast list decoding of a SPC node $o$ are exhibited in Algorithm \ref{Algorithm_SPC}. First, the absolute values in $\boldsymbol{\alpha}$ are sorted and then the least reliable bit which corresponds to the parity check constraint is estimated. Then $t=\min(L-1,N_o-1)$ path splits are implemented to judge $\tilde{\alpha}_i, i=1,...,t$. Whereafter, other bits in $\boldsymbol{\beta}$ can be directly judged by their corresponding LLRs in $\boldsymbol{\alpha}$. Note that after all other bits are obtained, the least reliable bit needs to be recomputed based on  (\ref{Equation_gamma}). Ultimately, $\mathbf{u}$ can be calculated through the inverse process of polar encoding and subsequently  $\mathbf{v}$ can be calculated through the inverse process of convolutional encoding.
\begin{algorithm}[h!]	
	\caption{Fast List Decoding of a SPC Node $o$} 
	\hspace*{0.02in} {\bf Input:} 
	$\boldsymbol{\alpha}[l], \mathbf{s}[l], PM_l, l=0,...,L-1$, $\mathbf{P}_{n-d_o},\mathbf{c}$\\
	\hspace*{0.02in} {\bf Output:} 
	$\boldsymbol{\beta}[l], \mathbf{s}[l], PM_l, \mathbf{\hat{v}}[l], \mathbf{\hat{u}}[l], l=0,...,L-1$
	\begin{algorithmic}[1]		
		\For{$l \leftarrow 0$ \textbf{to} $L-1$}
			\State $\hat{v}_0[l] \leftarrow 0$
			\State [$\hat{u}_0[l],\mathbf{s}[l]$] $\leftarrow$ conv1bitEnc($\hat{v}_0[l], \mathbf{s}[l],\mathbf{c}$)	
			\State $\boldsymbol{\beta}[l] \leftarrow \frac{1}{2}(1-$sgn($\boldsymbol{\alpha}[l]$) \texttt{// judgment}	
			\State $\gamma[l] \leftarrow (\sum_{j=0}^{N_o-1}\beta_j[l] + \hat{u}_0[l])\ \% \ 2$\texttt{//parity check} 			
	        \State [$|\boldsymbol{\tilde{\alpha}}[l]|,\mathbf{\tilde{I}}[l]$] $\leftarrow $ sort($|\boldsymbol{\alpha}[l]|$,`ascend')  \texttt{// sorting}
	        \State \texttt{// estimate the least reliable bit}		
			\If{$\gamma[l]=0$}
				\State $PM_l \leftarrow PM_l$
			\Else
				\State $PM_l \leftarrow PM_l+|\tilde{\alpha}_0[l]|$  
			\EndIf
		\EndFor
		\For{$i \leftarrow 1$ \textbf{to} $\min(L-1,N_o-1)$}  \texttt{// path split}
			\For{$l \leftarrow 0$ \textbf{to} $L-1$}
				\State $ll \leftarrow l+L$		
				\State [$\hat{u}'_0[l],\hat{u}'_0[ll]$] $\leftarrow$ [$\hat{u}_0[l],\hat{u}_0[l]$]
				\State [$\gamma'[l],\gamma'[ll]$]  $\leftarrow$ [$\gamma[l],\gamma[l]$]
				\State [$\boldsymbol{\beta'}[l],\boldsymbol{\beta'}[ll]$]  $\leftarrow$ [$\boldsymbol{\beta}[l],\boldsymbol{\beta}[l]$]
				\State [$\mathbf{s'}[l],\mathbf{s'}[ll]$] $\leftarrow$ [$\mathbf{s}[l],\mathbf{s}[l]$]
				\State [$\boldsymbol{\tilde{\alpha}'}[l],\boldsymbol{\tilde{\alpha}'}[ll] $ $\leftarrow$
				[$\boldsymbol{\tilde{\alpha}}[l],\boldsymbol{\tilde{\alpha}}[l]$]
				\State [$\mathbf{\tilde{I}'}[l],\mathbf{\tilde{I}'}[ll] $] $\leftarrow$
				[$\mathbf{\tilde{I}}[l],\mathbf{\tilde{I}}[l] $]
				\State [$\tilde{\beta}'_i[l],\tilde{\beta}'_i[ll]$] $\leftarrow$ [$0,1$]
				\State $PM'_l \leftarrow $ calcPM2($PM_l,\tilde{\alpha}'_i[l],\tilde{\beta}'_i[l], 1, \gamma'_i[l], \tilde{\alpha}_0$)	
				\State $PM'_{ll} \leftarrow $ calcPM2($PM_{l},\tilde{\alpha}'_i[ll],\tilde{\beta}'_i[ll], 1,\gamma'_i[ll], \tilde{\alpha}_0$)
			\EndFor	
			\State [$\mathbf{\widetilde{PM}},\mathbf{\tilde{l}}$] $\leftarrow$ sort([$PM'_0,...,PM'_{2L-1}$],`ascend')
			\For {$l \leftarrow 0$ \textbf{to} $L-1$}
				\State $PM_l$ $\leftarrow$ $\widetilde{PM}_l$ 
				\State $\mathbf{\tilde{I}}[l]$ $\leftarrow$ $\mathbf{\tilde{I}'}[\tilde{l}_l]$
				\State $\boldsymbol{\tilde{\alpha}}[l]$ $\leftarrow$ $\boldsymbol{\tilde{\alpha}'}[\tilde{l}_l]$
				\State $\mathbf{s}[l]$ $\leftarrow$ $\mathbf{s'}[\tilde{l}_l]$					
				\State $\boldsymbol{\beta}[l]$ $\leftarrow$ $\boldsymbol{\beta'}[\tilde{l}_l]$			
				\State $\beta_{\tilde{I}_i[l]}[l]$ $\leftarrow$ $\tilde{\beta}'_i[\tilde{l}_l]$	
				\State $\gamma[l]$ $\leftarrow$ $\gamma'[\tilde{l}_l]$	
				\State $\hat{u}_0[l]$ $\leftarrow$ $\hat{u}'_0[\tilde{l}_l]$	
			\EndFor			 
		\EndFor
		\For{$l \leftarrow 0$ \textbf{to} $L-1$}
			\State \texttt{// recompute the least reliable bit}	
			\State  $\beta_{\tilde{I}_0[l]}[l] \leftarrow (\sum_{j=1}^{N_o-1}\beta_{\tilde{I}_j[l]}[l] + \hat{u}_0[l]) \ \% \ 2 $
			\State \texttt{// inverse polar encoding}
			\State $\hat{\mathbf{u}}[l]	\leftarrow  \boldsymbol{\beta}[l] \cdot \mathbf{P}_{n-d_o}$	
			\For{$i \leftarrow 0$ \textbf{to} $N_o-1$}		
				\State \texttt{// inverse convolutional encoding}		
				\State [$\hat{v}_i[l],\mathbf{s}[l]$] $\leftarrow$ conv1bitInvEnc($\hat{u}_i[l], \mathbf{s}[l],\mathbf{c}$)				
			\EndFor
		\EndFor	
	\end{algorithmic}
	\label{Algorithm_SPC}
\end{algorithm}

\subsection{Time Steps}
Comparing list decoding with fast list decoding, we give the analysis of the required time steps for the four types of constituent nodes. We hypothesize that the $f$ operation (\ref{Equation_f}), the $g$ operation (\ref{Equation_g}) and one path split all consume one time step. Moreover, we ignore the time step of polar encoding and convolutional encoding. In addition, we do not take into account the resouce constriant, which means that all parallel operations can be carried out in one time step. In fact, these are reasonable hypotheses commonly used by scholars \cite{SC_delay,SCL_delay}.
\begin{itemize}
	\item Rate-0 Nodes: The original number of time steps is $2N_o-2$ consisting of $N_o-1$ $f$ operations and $N_o-1$ $g$ operations, while the fast list decoding can reduce it to $1$ because no $f$ and $g$ operation is needed and there is only one PM computation for a vector.
	\item Rate-1 Nodes: The original number of time steps is $3N_o-2$ consisting of $N_o-1$ $f$ operations, $N_o-1$ $g$ operations and $N_o$ path splits, while the fast list decoding can reduce it to $\min\{L-1,N_o\}$ with an effective algorithm for path split. 
	\item Rev Nodes: The original number of time steps is $2N_o-1$ consisting of $N_o-1$ $f$ operations, $N_o-1$ $g$ operations and one path split, while the fast list decoding can reduce it to $2$ with one path split and one PM computation for a vector. 
	\item SPC Nodes: The original number of time steps is $3N_o-3$ consisting of $N_o-1$ $f$ operations, $N_o-1$ $g$ operations and $N_o-1$ path splits, while the fast list decoding can reduce it to $\min\{L,N_o\}+1$ with one bit estimation for the least reliable bit, $\min\{L-1,N_o-1\}$ path splits and one bit recalculation.
\end{itemize}

Time steps of the four constituent nodes for list decoding and fast list decoding is listed in Table \ref{Table_TimeStep}. One thing that needs to be emphasized is that the performance of fast list decoding of Rate-0, Rate-1 and Rev nodes is exactly equivalent to that of list decoding, while the fast list decoding algorithm for SPC nodes is an approximation algorithm. Nevertheless, the performance loss is is so small as to be negligible as shown in Section \ref{Section_SimulationResults}.

\begin{table}[t!]		
	\caption{Time Steps of The Four Constituent Nodes for List Decoding and Fast List Decoding}		
	\begin{center} 
		\begin{tabular}{c|c|c}		
			\hline
			\hline
			Constituent Node & List Decoding & Fast List Decoding \\
			\hline
			Rate-0      & $2N_o-2$    &1                                  \\
			\hline
			Rate-1      & $3N_o-2$    &$\min\{L-1,N_o\}$                 \\
			\hline	
			Rev         & $2N_o-1$    &$2$                               \\
			\hline	
			SPC         & $3N_o-3$     & $\min\{L,N_o\}+1$               \\
			\hline
			\hline	
		\end{tabular}
		\label{Table_TimeStep}
	\end{center}
\end{table}

\section{Simulation Results}
\label{Section_SimulationResults}
In this section, we provide the simulation results of the fast list decoding unber different code lengths, code rates and list sizes. All PAC codes are obtained via RM rate-profiling using the rate-1 convolutional code generated by $\mathbf{c} = (1, 0, 1, 1, 0, 1, 1)$, just as the parameter setting in \cite{PAC_Arikan}. The encoded codeword $\mathbf{x}$ is modulated by binary phase shift keying (BPSK) modulation and transmitted over binary-input additive white Gaussian noise (BI-AWGN) channel. The BI-AWGN dispersion bound is obtained from \cite{DispersionBound}. We measure the FER performance against the $E_b/N_0$ in dB, where $E_b$ denotes the bit energy and $N_0$ denotes the power spectral density of the BI-AWGN noise. For each $E_b/N_0$, we ensure that at least 500 error frames occur to keep the results reliable. For the sake of concision, we refer to list decoding, fast list decoding with three types of constituent nodes (Rate-0 nodes, Rate-1 nodes and Rev nodes) and four types of constituent nodes (Rate-0 nodes, Rate-1 nodes, Rev nodes and SPC nodes) as `List', `Fast-List-Three' and `Fast-List-Four', respectively. 

\subsection{FER Performance}
\begin{figure}[t!]
	\centering
	\includegraphics[width=.45\textwidth]{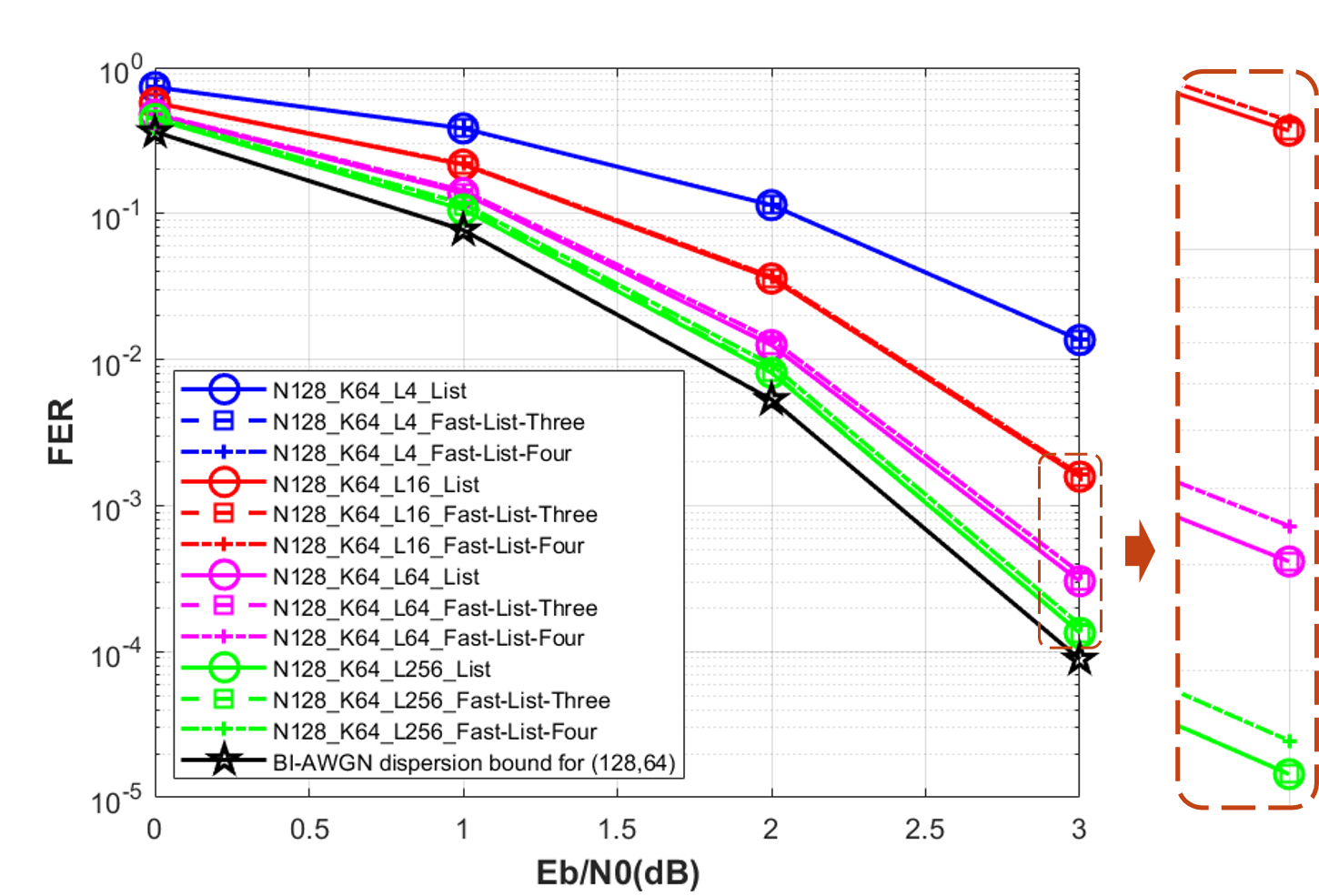}
	\caption{The FER performance comparison of List, Fast-List-Three and Fast-List-Four for $PAC(128,64)$ under different list sizes. The BI-AWGN dispersion bound for $(128,64)$ is also displayed.} 
	\label{Figure_FER1} 
\end{figure}
\begin{figure}[t!]
	\centering
	\includegraphics[width=.45\textwidth]{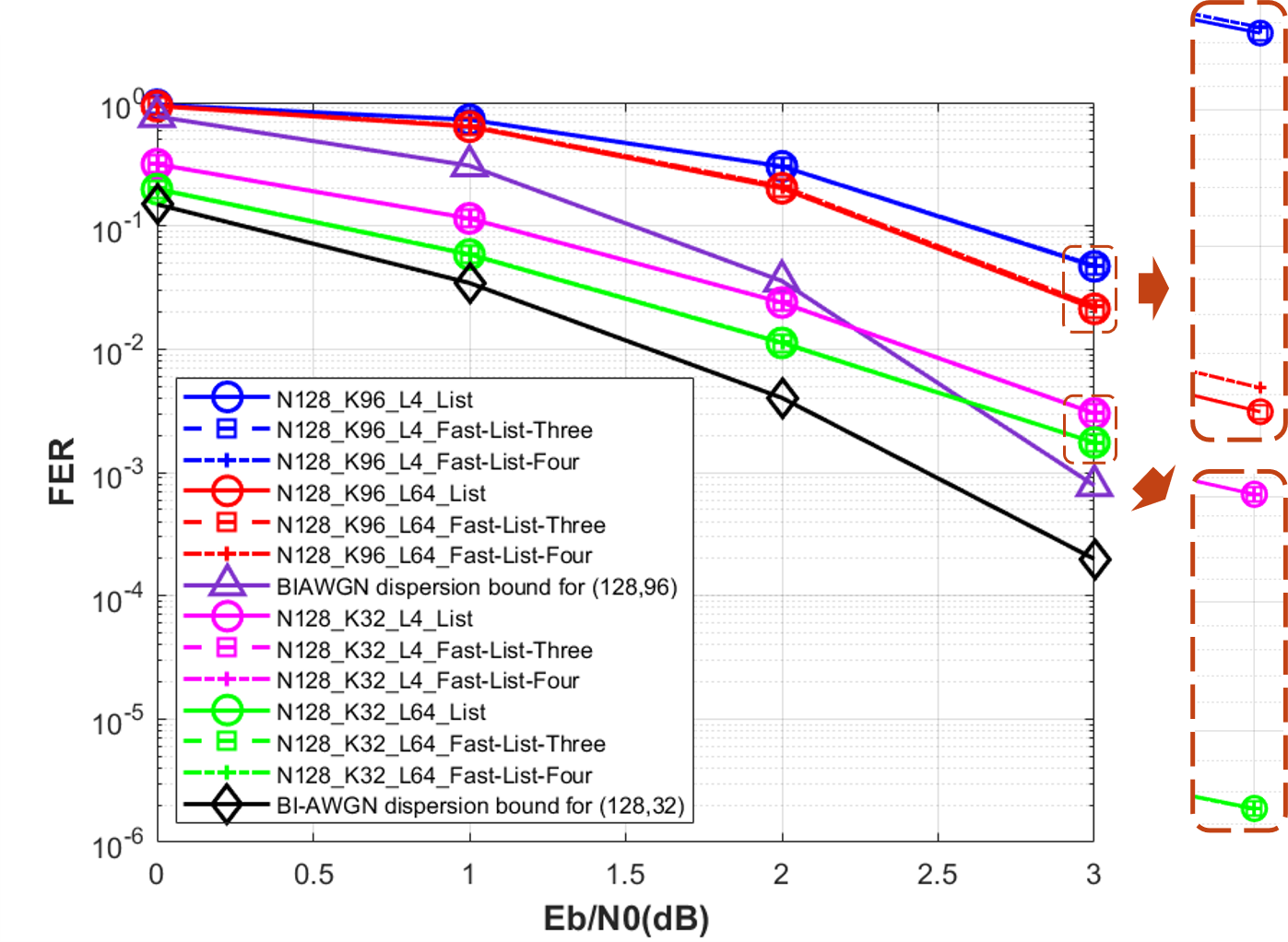}
	\caption{The FER performance comparison of List, Fast-List-Three and Fast-List-Four for $PAC(128,32)$ and $PAC(128,96)$ under different list sizes. The two BI-AWGN dispersion bounds are also displayed.} 
	\label{Figure_FER2} 
\end{figure}
\begin{figure}[t!]
	\centering
	\includegraphics[width=.45\textwidth]{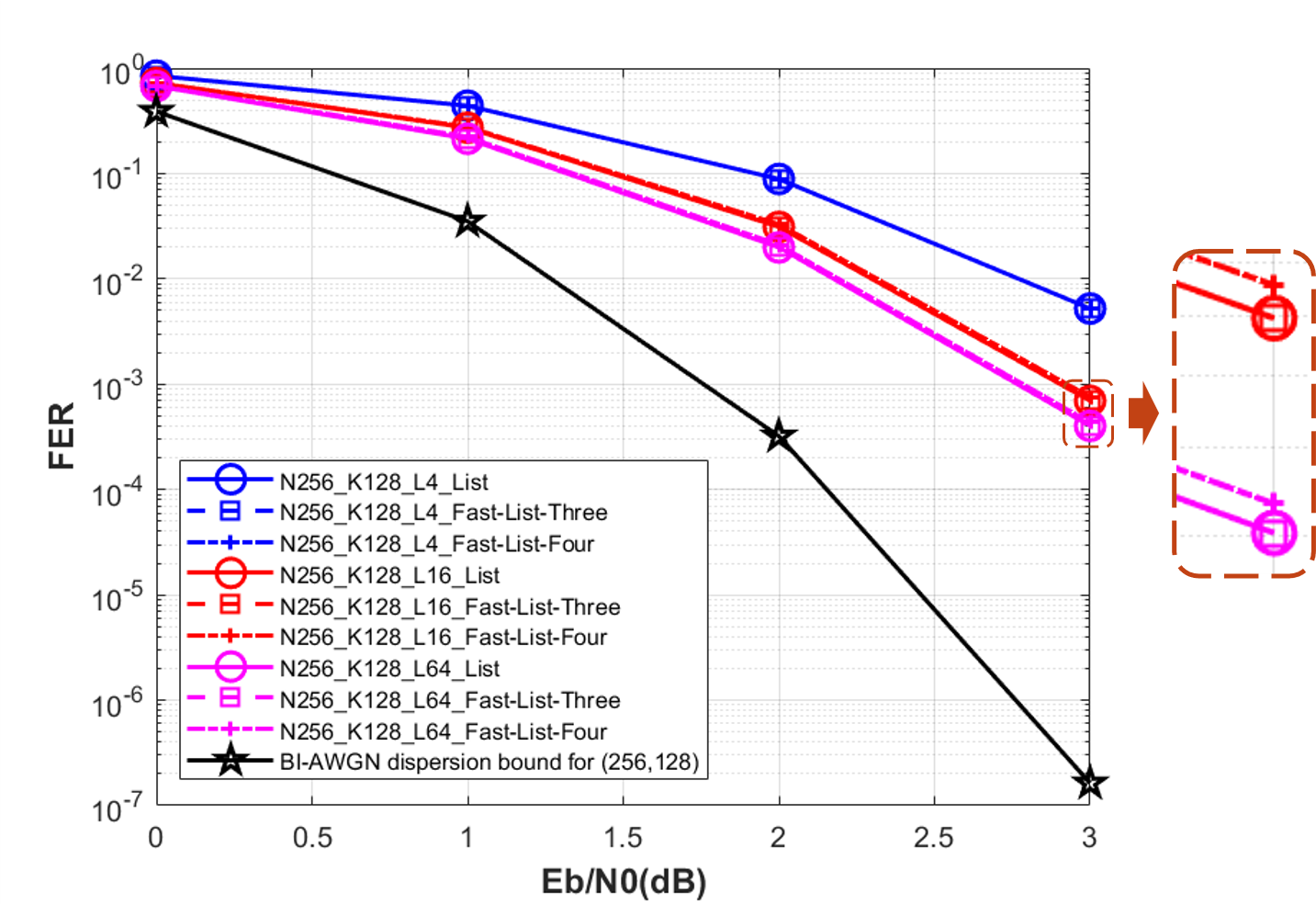}
	\caption{The FER performance comparison of List, Fast-List-Three and Fast-List-Four for $PAC(256,128)$ under different list sizes. The BI-AWGN dispersion bound for (256,128) is also displayed.} 
	\label{Figure_FER3} 
\end{figure}

At the beginning, we investigate the FER performance comparison of List, Fast-List-Three and Fast-List-Four for $PAC(128,64)$ under different list sizes, as shown in Fig. \ref{Figure_FER1}. The BI-AWGN dispersion bound $(128,64)$ is also displayed. According to the analysis in Section \ref{Section_ProposedAlgorithms}, Fast-List-Three is absolutely equivalent to List and Fast-List-Four is an approximation in theory. From Fig. \ref{Figure_FER1}, we can see that the simulation results exactly correspond to the theory. Fast-List-Three yields the same performance with List regardless of $L$ and approaches the BI-AWGN dispersion bound when $L=256$ with only 0.1 dB gap at FER $10^{-3}$. In contrast, Fast-List-Four incurs no performance degradation when $L=4$,  and causes neglectable performance loss when $L=16$, $L=64$ and $L=256$. 

Afterwards, we fix the code length and change the code rate. Fig. \ref{Figure_FER2} presents the FER performance comparison of List, Fast-List-Three and Fast-List-Four for $PAC(128,32)$ and $PAC(128,96)$ under different list sizes. The two BI-AWGN dispersion bounds are also displayed. We make sure that the performance of $PAC(128,32)$ and $PAC(128,96)$ has converged when $L=64$. It can be observed that all curves of Fast-List-Three and Fast-List-Four coincide with that of List for $PAC(128,32)$. It comes naturally to us that the low code rate leads to fewer SPC nodes, and therefore the performance loss of Fast-List-Four cannot be observed. For $PAC(128,96)$, Fast-List-Three achieves the equivalent performance while Fast-List-Four induces quite tiny difference. However, the convergent performance is inferior to the BI-AWGN bound with 0.5 dB gap at FER $10^{-2}$ for $PAC(128,32)$ and 0.8 dB gap at FER $10^{-1}$ for $PAC(128,96)$.

In the end, we fix the code  rate and adjust the code length. The FER performance comparison of List, Fast-List-Three and Fast-List-Four for $PAC(256,128)$ under different list sizes is depicted in Fig. \ref{Figure_FER3}. The BI-AWGN dispersion bound for (256,128) is also displayed. Also, Fast-List-Three achieves precisely the same performance as List. Although the number of SPC nodes increases as the code length increases, the performance degradation of Fast-List-Four remains negligible. However, the convergent performance of PAC codes with $L=64$ is so far from the dispersion bound. More advanced techniques to improve PAC codes with long code lengths are urgently needed. 

\subsection{Time Steps}
\begin{table}[t!]		
	\caption{Time Steps of List Decoding and Fast List Decoding Under Different Code Lengths, Code Rates and List Sizes.}		
	\begin{center} 
		\begin{tabular}{c|c|c|c|c}		
			\hline
			\hline
			\tabincell{c}{Code \\ Parameter }  & \tabincell{c}{$L$ }			
			& \tabincell{c}{List \\ Decoding }  & \tabincell{c}{Fast-List-Three\\ Decoding \\(Reduction)}  & \tabincell{c}{Fast-List-Four \\ Decoding \\ (Reduction)} \\ 
			\hline
			\multirow{2}{*}{\tabincell{c}{$PAC(128,32)$ \\ $R=1/4$ }} &4 &286 & 75 (73.78\%) & 72 (74.83\%)	\\			
			\cline{2-5}
			&64 & 286& 81 (71.68\%) & 78 (72.73\%)       \\
			\hline		
			\multirow{4}{*}{\tabincell{c}{$PAC(128,64)$ \\ $R=1/2$ }} &4 &318 & 143 (55.03\%) & 108 (66.04\%)\\	
			\cline{2-5}	
			&16 & 318 & 152 (52.2\%) & 132 (58.49\%)\\ 
			\cline{2-5}
			&64 & 318 & 152 (52.2\%) & 132 (58.49\%)\\ 
			\cline{2-5}
			&256 & 318 & 152 (52.2\%) & 132 (58.49\%)\\ 
			\hline
			\multirow{2}{*}{\tabincell{c}{$PAC(128,96)$ \\ $R=3/4$ }} &4 &350 & 145 (58.57\%) & 86 (75.43\%)  \\			
			\cline{2-5}
			&64 & 350 & 179 (48.85\%) & 150  (57.14\%) \\
			\hline	
			\multirow{4}{*}{$\tabincell{c}{$PAC(256,128)$ \\ $R=1/2$ }$} &4 & 638& 233 (63.48\%)& 163 (74.45\%)\\			
			\cline{2-5}
			&16 &638 &267 (58.15\%)&  215 (66.3\%) \\
			\cline{2-5}
			&64 & 638 & 268 (57.99\%) &   231 (63.79\%)   \\
			\hline
			\hline
		\end{tabular}
		\label{Table_TimeStep2}
	\end{center}
\end{table}	
To illustrate the speed advantage of the proposed fast list decoders, we summarize the time steps of list decoding and fast list decoding under different code lengths, code rates and list sizes in Table \ref{Table_TimeStep2}. From the table, we can observe that:
\begin{itemize}
	\item When $N$ and $R$ are fixed, a smaller $L$ will lead to a larger reduction for both Fast-List-Three and Fast-List-Four. The reason is that the fast list decoding of Rate-1 and SPC nodes has taken into account the effect of $L$, and smaller $L$ will lead to fewer time steps. When $L$ increases, the number of time steps will converge to a constant.
	\item When $N$ and $L$ are fixed, a lower $R$ tends to result in larger reduction for both Fast-List-Three and Fast-List-Four. The reason is that when the rate is lower, the number of Rate-0 and Rev nodes will become larger. The required time steps of Rate-0 and Rev nodes for fast list decoding are 1 and 2 respectively, which are so small that bigger gain will be brought about. 
	\item When $N$ and $L$ are fixed, Fast-List-Four has a greater advantage over Fast-List-Three if $R$ is higher. This is obvious because a higher $R$ will lead to more SPC nodes, and hence the fast list decoding algorithm of SPC nodes will show its superiority.
	\item When $R$ and $L$ are fixed, a bigger $N$ will lead to a larger reduction for both Fast-List-Three and Fast-List-Four. It is easy to think that when the $N$ increases, more constituent nodes will appear and therefore we can further accelerate decoding using our fast algorithms.
\end{itemize}

\section{Conclusion}
\label{Section_Conclusion}
In this paper, we propose fast list decoders for PAC codes. The fast scheme is mainly based on tree pruning technique which decodes constituent nodes on the top of the subtree without traversal of the binary tree. Fast list decoding algorithm of Rate-0, Rate-1 and Rev nodes are exactly equivalent to that of list decoding, while that of SPC nodes is an approximation. Consequently, the FER results demonstrate that fast list decoding with three types of constituent nodes incurs no performance degradation while the introduction of SPC nodes causes negligible performance loss. As the most remarkable superiority, the proposed decoders can significantly reduce the number of time steps so as to speed decoding in practice. 

As for the future direction, more constituent nodes may be found to further reduce the decoding latency. Besides, we have observed that the performance of PAC codes is still far from the dispersion bound when the code length increases to 256. More improvements about encoding and decoding are in demand to enhance the application universality of PAC codes.

\appendices
\section{Subroutines}
\label{Appendix_Subroutines}
Algorithm \ref{Algorithm_Subroutines} provides these subroutines required for algorithms in Section \ref{Section_ProposedAlgorithms}.

\begin{algorithm}[t]	
	\caption{Subroutines} 
	\label{Algorithm_Subroutines}
	\begin{algorithmic}[1]			
		\State \textbf{subroutine}	conv1bitEnc($v, \mathbf{s}, \mathbf{c}$)
		\State  $u \leftarrow v \cdot c_0$
		\For {$i \leftarrow 1$ \textbf{to} $m$}
			\If {$c_i=1$}
				\State $u \leftarrow u \oplus s_{i-1}$
			\EndIf
		\EndFor
		\State $\mathbf{s} \leftarrow [v, s_0, ..., s_{m-2}]$	
		\State \Return 	$u, \mathbf{s}$
		\State
		\State  \textbf{subroutine}	conv1bitInvEnc($u, \mathbf{s}, \mathbf{c}$)
		\State $v \leftarrow  0$
		\State $u_{temp} \leftarrow v \cdot c_0$
		\For {$i \leftarrow 1$ \textbf{to} $m$}
		\If {$c_i=1$}
		\State $u_{temp} \leftarrow u_{temp} \oplus s_{i-1}$
		\EndIf
		\EndFor
		\If {$u_{temp}=u$}
		\State $v \leftarrow 0$
		\Else
		\State $v \leftarrow 1$
		\EndIf
		\State $\mathbf{s} \leftarrow [v, s_0, ..., s_{m-2}]$	
		\State \Return 	$v, \mathbf{s}$
		\State
		\State  \textbf{subroutine}	calcPM($PM, \boldsymbol{\alpha}, \boldsymbol{\beta}, len$)		
		\For{$i \leftarrow 0 $ \textbf{to} $len-1$}
		\If {$\beta_i =\frac{1}{2}(1-$sgn($\alpha_i)$)} 
		\State $PM \leftarrow PM$
		\Else
		\State $PM \leftarrow PM+|\alpha_i|$
		\EndIf
		\EndFor	
		\State \Return 	$PM$
		\State
		\State  \textbf{subroutine}	calcPM2($PM, \boldsymbol{\alpha}, \boldsymbol{\beta}, len, \gamma, \tilde{\alpha}_0$)		
		\For{$i \leftarrow 0 $ \textbf{to} $len-1$}
		\If {$\beta_i =\frac{1}{2}(1-$sgn($\alpha_i)$)} 
		\State $PM \leftarrow PM$
		\Else
		\State $PM \leftarrow PM+|\alpha_i|+(1-2\gamma) |\tilde{\alpha}_0|$
		\EndIf
		\EndFor	
		\State \Return 	$PM$
	\end{algorithmic}
\end{algorithm}



\bibliographystyle{IEEEtran} 
\bibliography{references}

\end{document}